\documentclass[copyright,creativecommons]{eptcs}
\providecommand{\event}{TFPIE 2024} 

\usepackage{iftex}

\usepackage{hyperref}
\usepackage[capitalize]{cleveref}
\usepackage{xspace}

\usepackage[frozencache]{minted}     

\newmintinline[stella]{stella}{fontsize=\normalsize}
\newmintinline[kotlin]{kotlin}{fontsize=\normalsize}

\newcommand{\lama}{$\lambda\kern -.1667em\lower -.5ex\hbox{$a$}\kern -.1000em\lower .2ex\hbox{$\mathcal M$}\kern -.1000em\lower -.5ex\hbox{$a$}$\xspace}

\usepackage{todonotes}

\ifpdf
  \usepackage{underscore}         
  \usepackage[T1]{fontenc}        
\else
  \usepackage{breakurl}           
\fi

\title{Teaching Type Systems Implementation with \textsc{Stella}, an~Extensible Statically Typed Programming Language}
\author{
  Abdelrahman Abounegm \qquad\qquad
  Nikolai Kudasov \qquad\qquad
  Alexey Stepanov
\institute{Lab of Programming Languages and Compilers\\ Innopolis University\\ Innopolis, Tatarstan Republic, Russia}
\email{\quad a.abounegm@innopolis.university \quad\qquad n.kudasov@innopolis.ru \quad\qquad a.stepanov@innopolis.ru}
}
\def\titlerunning{Teaching Type Systems Implementation with \textsc{Stella}}
\def\authorrunning{A. Abounegm, N. Kudasov \& A. Stepanov}
\begin{document}
\maketitle

\begin{abstract}
  We report on a half-semester course focused around implementation of type systems
  in programming languages. The course assumes basics of classical compiler construction,
  in particular, the abstract syntax representation, the Visitor pattern, and parsing.
  The course is built around a language \textsc{Stella} with a minimalistic core and a set of small extensions,
  covering algebraic data types, references, exceptions, exhaustive pattern matching, subtyping,
  recursive types, universal polymorphism, and type reconstruction.
  Optionally, an implementation of an interpreter and a compiler is offered to the students.
  To facilitate fast development and variety of implementation languages we rely on the BNF Converter tool
  and provide templates for the students in multiple languages.
  Finally, we report some results of teaching based on students' achievements.
\end{abstract}


\section{Introduction}
\label{section:introduction}

Type systems constitute an important part of most modern programming languages,
from Java and C++ to Python and TypeScript to Scala and Haskell, to mention a few.
When using static types, programmers devote a significant part of
their interaction with the compiler or a static analysis tool working through the type errors.
Type-Driven Development (TyDD)~\cite{Brady2017} goes even further and suggests writing the types first
and then follow the types to produce an implementation.
Among popular modern programming languages, expressive type systems are becoming
more widespread. Two recent examples include Rust~\cite{MatsakisKlock2014}, a systems programming language
with ownership types, and Differentiable Swift~\cite{DifferentiableSwift}, a dialect of Swift
with differentiable types. Such languages incorporate quite elaborate type system features
that require solid understanding of the basics to be used efficiently.

Type systems are not only used in compilers, but also in proof assistants such
as Agda, Coq, and Isabelle/HOL. These usually rely on dependent type systems,
such as Martin-L\"of Type Theory~\cite{MartinLof1984} or Calculus of Constructions~\cite{CoquandHuet1988}.
Using proof assistants for mathematics or program verification is somewhat similar in practice
to TyDD in programming languages: in both cases the user of the system spends
a significant amount of time dealing with type errors.

Although linters and good error messages help with type errors,
it is important for the users to understand and be able to follow
typing rules to productively resolve the issues and utilize type checker
as a helping tool, instead of seeing it as a barrier.
Although courses in statically typed programming languages help teach students
some of the mechanics behind type checking, we believe a first-hand experience
in the implementation of a type system deepens that understanding
while also allowing students to apply some of those types in the implementation.
Unfortunately, existing compilers construction courses appear to either
provide an overview of the compiler pipeline, resort to a simplified semantic analysis phase,
focus on the code generation phase, or work with a low-level intermediate representation
such as LLVM IR or Java bytecode.


On the other hand, courses that focus on type systems often offer
implementation of a variant of typed lambda calculus, which appears to
disengage some students who are used to Java-style or Python-style syntax
of programming languages and who do not have sufficient experience with functional programming.
In particular, in a previous iteration of our course,
many students found it quite difficult and unintuitive to program or even read programs in lambda calculus,
whereas rewriting a lambda expression in Python using explicit function definitions
helped them navigate the operational semantics of lambda calculus.


While lambda calculus is too raw, any modern programming language is too complex.
Indeed, even the core of most modern languages is too complex to give as an implementation exercise
for students. The Haskell programming language, and in particular its implementation
in the Glasgow Haskell Compiler (GHC), comes close to a language with a small core:
it employs a system of language extensions that allows to enrich the language per module.
In this work, we are inspired by this approach and design a language with a very small fixed core
that is easy to implement for students, and provide various extensions for further developments.

To be able to focus on the type checking, it is not enough to assume that students
have already mastered the lexical and syntactical analyzers. Implementing those
with proper types for the abstract syntax tree from scratch still requires significant time.
Luckily, at least for simple languages, there is a good selection of tools that
can do that automatically, such as XText~\cite{EysholdtBehrens2010} and BNF Converter~\cite{BNFC}.
In our course, we rely heavily on BNF Converter to offer templates in multiple languages
and encourage diversity of implementations in student submissions.


\subsection{Related Work}
\label{section:related}


The idea of designing a programming language specifically for teaching compiler construction is not new.
One of the first such languages was \texttt{MINIPASCAL}, a simplified version of Pascal programming language, introduced by Appelbe~\cite{Appelbe1979}.

More recently, Berezun and Boulytchev~\cite{BerezunBoulytchev2022} reported on the
use of the programming language \lama{}\footnote{\url{https://github.com/PLTools/Lama}},
developed by JetBrains Research for educational purposes, in their introductory course on compilers.
They asked students to implement a compiler of \lama{} in itself, a procedural
language with first-class functions that is untyped (meaning no \emph{static} type checking is performed).
It appears that the choice to make the language untyped was to afford simpler implementations.
Thus, Berezun and Boulytchev's course focuses on parsing, basic semantic analysis,
and code generation. In contrast, our course focuses on type systems and puts semantics first.

Aiken~\cite{Aiken1996} introduced the Cool language, which is largely inspired by Java and features a static type system
and automatic memory management. Cool's type system is nominal, without support for generics,
anonymous objects, or higher-order functions, and thus does not present significant challenges
in the implementation of type checking compared to the structural type system featured in our course.




Although basic typechecking can be covered in regular compilers courses,
some courses still specialize specifically on type systems and type checking.
Ortin, Zapico, and Cueva~\cite{OrtinZapicoCueva2007} teach design
patterns helpful for implementing a type checker in object-oriented languages.


L{\"u}bke, Fuger, Bahnsen, Billerbeck, and Schupp~\cite{LubkeFugerBahnsenBillerbeckSchupp2023}
show how to automate exams and assignments for functional programming that include proofs
traditionally performed on paper. With \textsc{Stella}'s system of extensions,
we have the basis for similar automation for our assignments, but we keep
advanced automation for future work.


\subsection{Contribution}

In this paper, we report on a half-semester course on the implementation
of type systems. More specifically:

\begin{enumerate}
  \item In \cref{section:stella}, we overview the \textsc{Stella} language,
  consisting of a minimalistic purely functional core and a set of extensions,
  matching multiple topics covered by Pierce in his book~\cite{TaPL}.
  \item In \cref{section:templates}, we specify our approach to template solutions,
  allowing students to get started with implementations in any of a multitude of languages,
  including C++, Java, Kotlin, OCaml, TypeScript, Swift, Rust, Go, and Python.
  \item In \cref{section:course}, we outline the course structure and identify
  both practical and theoretical tasks for the students.
  Here, we also evaluate the course based on students' performance.
\end{enumerate}

\section{Overview of the \textsc{Stella} Language}
\label{section:stella}

We have designed and implemented the
\textsc{Stella}\footnote{\textbf{S}tatically \textbf{T}yped \textbf{E}xtensible \textbf{L}anguage for \textbf{L}earning \textbf{A}dvanced compiler construction (type systems)}
language specifically for the purposes of teaching type systems implementation
following Pierce's book~\cite{TaPL}. In this section, we highlight the main design decisions
behind the \textsc{Stella} language and outline its main features.

\textsc{Stella} is a statically-typed expression-based language.
It consists of a minimalistic core language and a set of extensions.
Making language extensions explicit is advantageous in several ways:
\begin{enumerate}
  \item Students can quickly learn the core, explicitly see a list of features
        used in every example program, and know where to look for corresponding documentation;
  \item Test programs explicitly specify the language extensions used,
        making it easy to understand which features (or combinations of features)
        are supported by a student's implementation;
  \item Students can ignore extension pragmas, implicitly supporting
        any given set of extensions, simplifying implementations;
  \item More extensions can be added to \textsc{Stella} in the future.
\end{enumerate}

The canonical implementation of Stella supports all extensions. Templates
provided to the students also support all extensions, meaning that the abstract
syntax types are provided for the full language. Although this slightly complicates
the types for the core language (e.g. students have to deal with a list of function
arguments instead of a single argument), we find that in practice students do not have
any major issues with this.


A program in \textsc{Stella} always consists of a single file since a module
system is out of the scope of the course as it adds unnecessary complexity while
not adding much to the typechecking part of the compiler.

Below, we describe the core part of the language and the main extensions featured in the course.

\subsection{\textsc{Stella} Core}

The core language of \textsc{Stella} is essentially a simply-typed functional programming language,
based on Programming Computable Functions (PCF)~\cite{Plotkin1977LCFCA},
except having a syntax inspired particularly by Rust~\cite{MatsakisKlock2014},
and reasonably understandable to many programmers.
In fact, one of the main motivations for \textsc{Stella} was the fact that many students
struggled with the \emph{syntax} of $\lambda$-calculus but many examples were better understood
(and students came up more easily with their own examples) when moving to a more familiar syntax
of pseudocode or Python, Java, TypeScript.

\textsc{Stella} Core also corresponds to simply-typed lambda calculus~\cite[\S 9]{TaPL} with
two base types (natural numbers and booleans)~\cite[\S 11.1]{TaPL}.

Importantly, \textsc{Stella} Core supports first class functions.
We justify support for higher-order functions from the start by asking students
to only implement a typechecker, leaving the overview of possible implementations
for a proper compiler for a functional language to a later part of the course.
Indeed, while a compiler might need to deal explicitly with closures and renaming of bound variables,
in a typechecker (without extensions like universal polymorphism or dependent types)
handling of scopes is straightforward and is not affected by the presence of higher-order functions.
At the same time, allowing higher-order functions allows us to immediately
introduce more sophisticated test programs.

\begin{figure}
  \begin{minted}[linenos]{stella}
// sample program in Stella Core
language core;

fn increment_twice(n : Nat) -> Nat {
  return succ(succ(n))
}

fn main(n : Nat) -> Nat {
  return increment_twice(succ(n))
}
\end{minted}
  \caption{Simple program in \textsc{Stella} Core.}
  \label{figure:stella-example}
\end{figure}

A sample program in \textsc{Stella} Core is given in \cref{figure:stella-example}:

\begin{enumerate}
  \item Line 1 is a comment line; all comments in \textsc{Stella} start with \stella{//}; there are no multiline comments in \textsc{Stella};
  \item Line 2 specifies that we are using just \textsc{Stella} Core (without any extensions); it is mandatory to specify the language in the first line of a Stella program;
  \item Line 4 declares a function \stella{increment_twice} with a single argument \stella{n} of type \stella{Nat} and return type \stella{Nat};
    in \textsc{Stella} Core, all functions are single-argument functions;
  \item In Line 5, \stella{return ...;} is a mandatory part of every function;
    in \textsc{Stella} Core each function can only return some expression;
    there are no assignments, operators, or other statements possible;
  \item In Line 5, \stella{succ(n)} is an expression meaning $n + 1$ (the successor of $n$);
  \item Line 8 declares a function \stella{main} with a single argument \stella{n} of type \stella{Nat} and return type \stella{Nat};
    in a Stella program there must always be a \stella{main} function declared at the top-level;
    the argument type and return type of \stella{main} can be specified to be any valid types;
  \item In Line 9, \stella{increment_twice(succ(n))} is an expression meaning ``call \stella{increment_twice} with the argument \stella{succ(n)}''.
\end{enumerate}


\subsection{Extensions}

\textsc{Stella} supports a number of language extensions,
which can be enabled with an \stella{extend with} pragma,
containing a list of extension names separated by commas.
Each extension may add new syntax, typing rules, or other capabilities to the language.
Following Pierce~\cite{TaPL}, we separate \emph{Simple Extensions} from
some more advanced type system extensions.
However, we also separate
\emph{Syntactic Sugar and Derived Forms} and \emph{Base Types} into standalone categories.

In general, each extension is supposed to be small enough to be implemented either
as a single exercise or at most as a standalone assignment. For example,
each base type has its own extension. Some extensions have simple and generalized versions.
For instance, \stella{#tuples} generalizes \stella{#pairs} by allowing arbitrary size and \stella{#variants}
generalizes \stella{#sum-types}.

\subsubsection{Syntactic Sugar and Derived Forms}

This category contains language extensions that \emph{may} be implemented as
derived forms, reducing them to other features of \textsc{Stella}.
However, this is not enforced and students are allowed to implement
these extensions as standalone features.
Some notable extensions in this category include
\href{https://fizruk.github.io/stella/site/extensions/syntax/#let-bindings}{\texttt{let}-bindings},
\href{https://fizruk.github.io/stella/site/extensions/syntax/#nested-function-declarations}{nested function declarations},
\href{https://fizruk.github.io/stella/site/extensions/syntax/#multiparameter-functions}{multiparameter functions},
\href{https://fizruk.github.io/stella/site/extensions/syntax/#automatic-currying}{automatic currying}, and
\href{https://fizruk.github.io/stella/site/extensions/simple-types/#type-ascriptions}{type ascriptions}.

A simple, but important extension is \href{https://fizruk.github.io/stella/site/extensions/syntax/#sequencing}{sequencing}.
Coupled with effectful expressions (such as mutable references), this enables imperative programming features.

\subsubsection{Nested Pattern Matching}

Some extensions (such as sum types or variants) naturally come equipped with pattern matching constructions.
However, by default they do not allow nested patterns, to simplify implementation and first example.
\emph{Nested} pattern matching is available via the \stella{#structural-patterns} extension.
This extension allows nesting and combining patterns that are enabled by other extensions.
Nested patterns complicate \emph{pattern-match coverage checking} (also known as \emph{exhaustiveness checking})
even just in presence of both tuples and variants and although it is a well-studied problem~\cite{Krishnaswami2009,Maranget2007,Sestoft1996}
we leave the implementation of the exhaustiveness checker out of the scope of our course
and only ask for exhaustiveness checks without nested patterns.

\subsubsection{Simple Types}

The simple types correspond directly to Pierce's \emph{Simple Extensions}~\cite[\S 11.2, \S 11.6--11.10, \S 11.12]{TaPL}
and include the \stella{Unit} type, pairs, tuples, records, sum types, variants, and (built-in) lists.

Importantly, all these types are \emph{structural} rather than \emph{nominal}~\cite[\S 19.3]{TaPL}.

\subsubsection{References}

\textsc{Stella} introduces references following Pierce~\cite[\S 13]{TaPL}.
Without any ownership types, this does not introduce any complications in the typechecking.

\subsubsection{Exceptions}

Following Pierce~\cite[\S 14]{TaPL}, we introduce support for different treatment of exceptions in \textsc{Stella}:

\begin{enumerate}
  \item First, simple (unrecoverable) errors~\cite[\S 14.1]{TaPL} are supported
  in the \stella{panic!} expression, enabled through \stella{#panic} extension;
  \item To throw and catch exceptions carrying values~\cite[\S 14.3]{TaPL},
  the extension \stella{#exceptions} is used. To specify the type of values in exceptions,
  the user has a choice:
  \begin{enumerate}
    \item One option allows user to fix the type of values carried by exceptions.
    This is achieved with the \stella{#exception-type-declaration} extension and requires
    a specification of exception type:
    \begin{minted}{stella}
// use error codes for exceptions
exception type = Nat
// use a fixed variant type for exceptions
exception type = <| error_code : Nat, good : Bool |>
    \end{minted}
    \cref{figure:stella-exceptions} provides an example of a complete \textsc{Stella} program
    that features exceptions.

    \item Another option is to use open variant type for exceptions (OCaml-style).
    This is achieved with the \stella{#open-variant-exceptions} extension and allows
    adding variants to the exception as needed. The following two declarations
    are equivalent to the explicit variant type above, except, with open variant,
    it can also be extended with more variants later:
\begin{minted}{stella}
exception variant error_code : Nat
exception variant good : Bool
\end{minted}
  \end{enumerate}
\end{enumerate}

\begin{figure}
\begin{minted}{stella}
language core;

extend with #exceptions, #exception-type-declaration;

exception type = Nat

fn fail(n : Nat) -> Bool {
  return throw(succ(0))
}

fn main(n : Nat) -> Bool {
  return try { fail(n) } with { false }
}
\end{minted}
\caption{A \textsc{Stella} program featuring exceptions.}
\label{figure:stella-exceptions}
\end{figure}

At the moment, \textsc{Stella} does not support any annotations that would
specify the possible exceptions thrown by a function.

\subsubsection{Subtyping}

\textsc{Stella} supports structural subtyping, following Pierce~\cite[\S 15]{TaPL}.
The subtyping mechanism is enabled with the
\stella{#structural-subtyping} extension and interacts with other enabled extensions,
such as \stella{#records} and \stella{#variants}.

In \cref{figure:stella-subtyping-records} on line 10, the function \stella{getX} is
applied to a record of type \stella{{x : Nat, y : Nat}}. Since this type is a
subtype of \stella{{x : Nat}}, it is accepted when structural subtyping is enabled.
Similarly, in \cref{figure:stella-subtyping-variants} on line 17, the function \stella{inc} is
applied to a value of variant type \stella{<| value : Nat |>}. Since this type is a
subtype of \stella{<| value : Nat, failure : Unit |>}, it is accepted when structural subtyping is enabled.

Subtyping for function types is enabled automatically and presents a
challenge for students, since it requires them to properly understand
the idea of covariant and contravariant subtyping. We find this also to be one
of the most valuable learning points for the students, since even without typing,
these concepts are important in programming.


\begin{figure}
\begin{minted}[linenos]{stella}
  language core;

  extend with #records, #structural-subtyping;

  fn getX(r : {x : Nat}) -> Nat {
    return r.x
  }

  fn main(n : Nat) -> Nat {
    return getX({x = n, y = n});
  }
\end{minted}
\caption{Sample \textsc{Stella} program with structural subtyping and records.}
\label{figure:stella-subtyping-records}
\end{figure}

\begin{figure}
\begin{minted}[linenos]{stella}
  language core;

  extend with #variants, #structural-subtyping;

  fn inc(r : <| value : Nat, failure : Unit |>) -> Nat {
    return match r {
        <| value = n |> => succ(n)
      | <| failure = _ |> => 0
    }
  }

  fn just(n : Nat) -> <| value : Nat |> {
    return <| value = n |>
  }

  fn main(n : Nat) -> Top {
    return inc(just(n));
  }
\end{minted}
\caption{Sample \textsc{Stella} program with structural subtyping and variants.}
\label{figure:stella-subtyping-variants}
\end{figure}

Additional extensions introduce \stella{Top} and \stella{Bot} (short for ``Bottom'') types as well.
The \stella{Top} type is the supertype of all types,
while the \stella{Bot} type is the subtype of all types.
The \stella{Top} type is not very useful on its own, so is normally used
in combination with the \stella{#type-cast} extension, which adds the
\stella{<expression> cast as <type>} expression syntax.
This operator performs downcasting of the expression to the specified type,
which only works if the expression was already of a supertype of the specified type.

\textsc{Stella} does not currently support intersection or union types~\cite[\S 15.7]{TaPL}.

\subsubsection{Universal Polymorphism}

Universal types~\cite[\S 23]{TaPL} are implemented in Stella in the form of generic functions that
accept type arguments. This is enabled by the \stella{#universal-types} extension.
To declare a generic function, one needs to add the \stella{generic} keyword
before \stella{fn} and add a type parameter list after the function name.
The type parameters are a comma-separated list of type variables enclosed in
square brackets. When invoking a generic function, the type arguments must be
provided in square brackets after the function name.

\cref{figure:stella-universal-polymorphism} shows a sample program with universal polymorphism:
\begin{enumerate}
  \item In Line~5, we declare function \stella{id} that is parametrized by type \stella{T};
  \item In Line~10, we apply the function \stella{id} to an argument,
  explicitly instantiating the type parameter to \stella{Nat}.
\end{enumerate}

\cref{figure:stella-universal-polymorphism-lambda} features a universally polymorphic anonymous function:
\begin{enumerate}
  \item In Line~5, we declare the function \stella{const} that returns a parametrically polymorphic function
  of type \stella{forall Y. fn(Y) -> X};
  \item In Line~6, we construct an anonymous function of said type;
  \item In Line~10, we first instantiate the type parameter of \stella{const} to \stella{Nat},
  then pass \stella{x} argument to get a parametrically polymorphic function
  of type \stella{forall Y. fn(Y) -> Nat} as a result. Finally, we instantiate \stella{Y} with \stella{Bool}
  and pass \stella{false} as an argument.
\end{enumerate}

\begin{figure}
\begin{minted}[linenos]{stella}
language core;

extend with #universal-types;

generic fn id[T](x : T) -> T {
  return x
}

fn main(x : Nat) -> Nat {
  return id[Nat](x)
}
\end{minted}
\caption{Sample \textsc{Stella} program with universal polymorphism.}
\label{figure:stella-universal-polymorphism}
\end{figure}

\begin{figure}
\begin{minted}[linenos]{stella}
language core;

extend with #universal-types;

generic fn const[X](x : X) -> forall Y. fn(Y) -> X {
  return generic [Y] fn(y : Y) { return x }
}

fn main(x : Nat) -> Nat {
  return const[Nat](x)[Bool](false)
}
\end{minted}
\caption{Sample \textsc{Stella} program with a universally polymorphic anonymous function.}
\label{figure:stella-universal-polymorphism-lambda}
\end{figure}

Currently, \textsc{Stella} does not support any form of bounded quantification~\cite[\S 26]{TaPL},
as a half semester course does not have enough room to explore this important topic in details.
However, the topic is briefly mentioned in the lectures. In particular,
the undecidability of type reconstruction in presence of bounded quantification~\cite[\S 28.5.5]{TaPL} is discussed.

The \stella{#universal-types} extension enables unrestricted impredicative universal types
as in System~F~\cite{Girard1986,Reynolds1974}. The following example demonstrates
impredicativity through self-application of \stella{f} to itself:

\begin{minted}{stella}
generic fn self_app[X](f : forall X . fn(X) -> X) -> forall X . fn(X) -> X {
  return f[forall X . fn(X) -> X](f)
}
\end{minted}

In the future, we aim to support various kinds of universal polymorphism in \textsc{Stella},
at least allowing for Hindley-Milner-style, predicative, and bounded quantification (parametric polymorphism with subtyping)~\cite[\S 26]{TaPL}.

\subsubsection{Recursive Types}

Recursive types~\cite[\S 20]{TaPL} (specifically, \emph{iso-recursive types}) are technically implemented in \textsc{Stella},
but not used in the course. The reason for that is that most programming languages rely on nominal typing
to provide recursion in types. Thus, in the course, we discuss recursive types in theoretical materials,
but do not offer implementation of this part of \textsc{Stella}.

\subsection{Accessibility: Documentation and Interactive Playground}

To provide the students with a convenient way to learn the language and its
extensions, we have created a website with documentation and an interactive
playground. The website is available at \url{https://fizruk.github.io/stella/}.

\begin{figure}
  \centering
  \includegraphics[width=\textwidth]{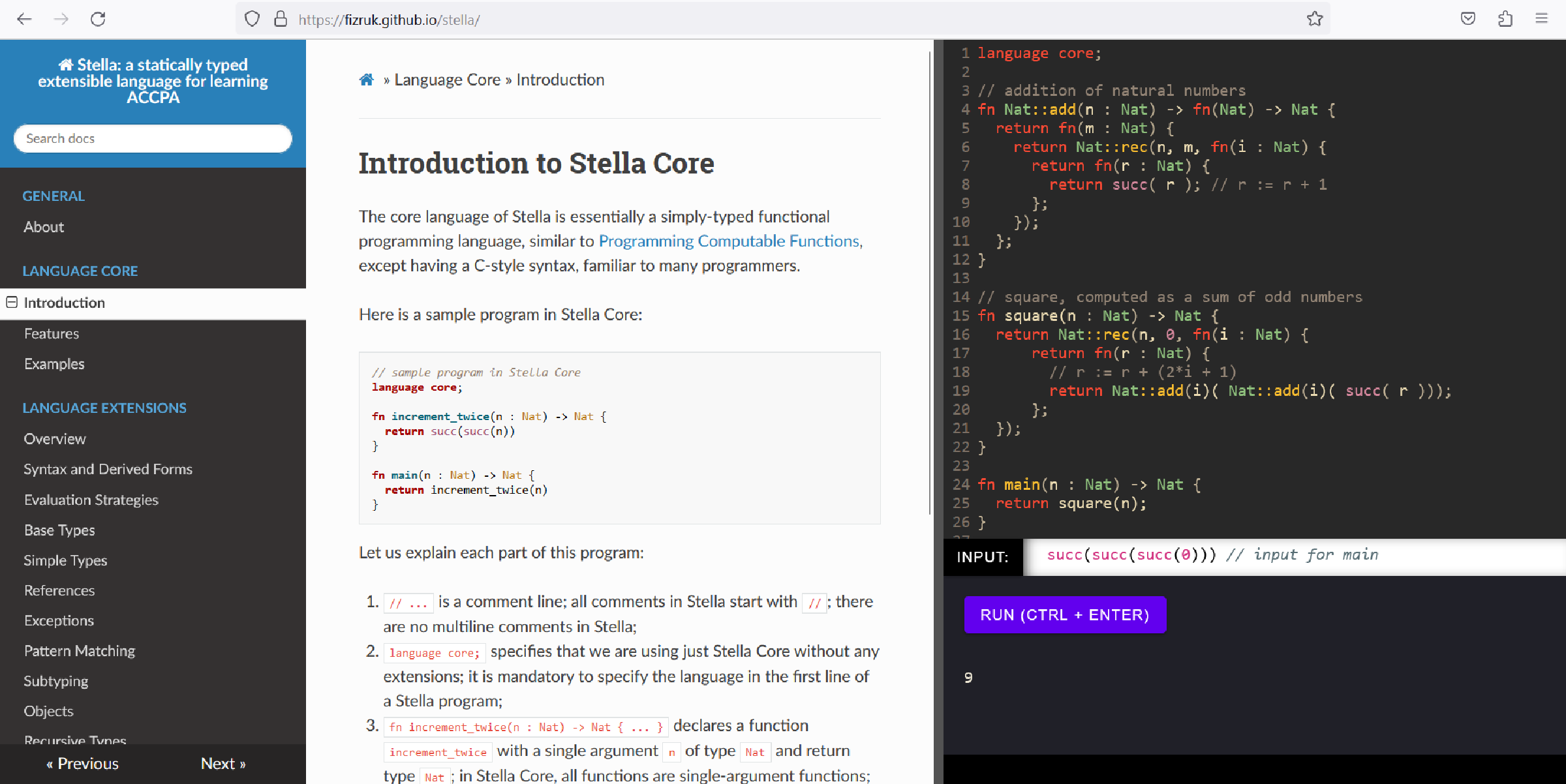}
  \caption{Stella documentation and playground.}
  \label{fig:stella-docs}
\end{figure}

The documentation is written in Markdown and is compiled to HTML using MkDocs~\footnote{\url{https://www.mkdocs.org}}.
The browser version of the Stella interpreter is compiled using GHCJS~\cite{GHCJS} and
Miso framework\footnote{\url{https://github.com/dmjio/miso}}, which allow us to compile Haskell code to JavaScript and run it in the browser.
The playground consists of a CodeMirror\footnote{\url{https://codemirror.net/}} code editor
with custom syntax highlighting using Highlight.js\footnote{\url{https://highlightjs.org/}}.
The playground includes an input field which acts as the \stella{main} function argument,
and an output field which displays the result of the \stella{main} function
or the compilation errors (if any).

Additionally, a VS Code extension is available for Stella, which provides
syntax highlighting and snippets for the language\footnote{\url{https://github.com/IU-ACCPA-2023/vscode-stella/}}.

\section{Implementation Templates and Typechecker Structure}
\label{section:templates}

By design, students are free to select any implementation language in the course.
To facilitate with implementations in many languages and avoid eating up the time
for parsing, abstract syntax implementation, and pretty-printing, we provide
students with an EBNF grammar, suitable for the BNF Converter tool~\cite{BNFC}.
See \cref{section:course} for the course setup, where we explain the prerequisites
allowing us to have a fast setup and skip a major part of the frontend for the student implementation.
The tool allows us to generate a lot of structure in many languages, including
Haskell, OCaml, Java, and C++. To reach even more languages, we have leveraged
the Java ANTLR backend of BNFC to produce ANTLR grammar with Java code snippets.
We then manually removed the Java code from the ANTLR grammar and added extra annotations to provide
a language-agnostic ANTLR file\footnote{see, for instance, \url{https://github.com/IU-ACCPA-2023/stella-implementation-in-swift/blob/main/Sources/stella-implementation-in-swift/Stella/stellaParser.g4}}.
Using ANTLR backends for other languages and writing down some definitions and conversions,
we have been able to get decent templates for more languages, including Rust, Swift, and TypeScript.







\subsection{A Zoo of Implementation Languages}

For convenience of students, we have prepared project templates\footnote{\url{https://github.com/IU-ACCPA-2023}} in several programming languages.
The templates allow to immediately proceed with the solution of the course objectives (implementation of typecheckers),
without spending extra time on the implementation of lexer and parser, definition of the abstract syntax, or implementing a pretty-printer.
The templates were prepared for the following languages: C++, Java, OCaml, Python, Rust, Go.
During the course, at the request of the students, and with their help, the following languages were added to the list: Kotlin, Swift, TypeScript, Haskell.

In the course, each student selects an implementation language, creates a copy of a suitable template, completes the coding exercises, and submits the solution via GitHub Classroom
(by pushing to the remote repository created for a particular assignment).

The templates were initially designed with the following structure, mostly provided by BNFC:
\begin{enumerate}
  \item A grammar file (Labelled BNF grammar) describing the \textsc{Stella} language;
  \item Lexer and parser generated by the BNFC tool based on the grammar file;
  \item The types (e.g. classes) for the abstract syntax of \textsc{Stella};
  \item A generated skeleton for recursive traversal of the abstract syntax;
  \item The main project file, which contains the functionality to read \textsc{Stella} source and convert it to AST;
  \item A \texttt{Makefile} to compile and run the project;
  \item Some tests to verify the success of building and running the project.
\end{enumerate}

At the start of the course, we found out that many students are not comfortable enough with \texttt{Make},
so we have accommodated for more IDE-friendly setups, relying on language-specific build systems, such as
CMake for C++, Dune for OCaml, and Maven for Java.

The Kotlin template is based on the Java pattern, meaning that the parser is implemented using Java classes.
However, the interpreter code is written in Kotlin and the build system has been replaced by Gradle.

The ANTLR grammar produced by BNFC for the Java backend was used as the basis to accommodate templates for more languages.
It was manually rewritten into a language-agnostic form (removing Java code, adding labels for rules, and slightly modifying the structure of the grammar).
We then leveraged ANTLR to generate the \emph{context types} for concrete syntax tree
and manually added types for abstract syntax together with corresponding conversion functions for each of the extra languages.

\section{Course Structure}
\label{section:course}

In this section, we discuss the organization of the course, its materials, the instruments used, and the results of the course.

The course consists of two lectures and two lab sessions each week for a total of 15 lectures and 15 labs over the course of eight weeks.
There are 4 coding and 3 theoretical assignments in the course, as well as an optional oral exam at the end of the course.
The first 6 weeks of the course cover type systems following Pierce~\cite[\S 8--23]{TaPL}, and the last 2 weeks cover
implementation details for lazy functional languages, following Peyton Jones~\cite[\S 1--5]{PeytonJones1992}.

The course is focused on the study of type systems, their properties and implementation.
For the theoretical part of the course, students are expected to learn about type safety,
canonical representation, normalization, and be able to prove these properties for
simple type systems. The practical part consists of implementing a type checker
for \textsc{Stella} with a subset of extensions.
Since many of the students taking the course already have experience with statically typed languages
like Java, C\#, C++, Scala, Kotlin, Go, Dart, and TypeScript, an implicit additional objection
of the course is to help students better understand and utilize type system features
in those languages, as well as appreciate the diversity of type systems and practical choices
made by designers of those systems.

\subsection{Prerequisites}

For the most part, the course emphasizes a detailed study of a specific component of compilers - the type checker.
Our course immediately follows another half-semester introductory course on compiler construction,
during which students developed the skills to construct their own compiler from the ground up.
Thus, entering our course, students are expected to have a general understanding of the structure of a compiler and be especially comfortable dealing with the abstract syntax.
Familiarity with compilers from textbooks by Wirth~\cite{Wirth1996}, Muchnick~\cite{Muchnick1998}, and Appel~\cite{Appel2004ML,AppelGinsburg2004C,AppelPalsberg2003Java}
is assumed.
Therefore, our course omits these aspects and focuses on more in-depth semantic analysis, type checking~\cite[\S 8--23]{TaPL},
and runtime for lazy functional languages~\cite{PeytonJones1992}.

The course additionally assumes familiarity with some statically typed programming languages.
In particular, we assume students to be able to write object-oriented code in C++ or Java,
and be able to write functional code in Haskell, OCaml, Scala, or a similar language.

The course load is approximately 15 hours per week. This includes two lectures, two laboratories, and an estimated time for homework. In addition to this course, students are enrolled in two other courses, which are also organized in a block scheme.

\subsection{Blocks}

The course was organized into five blocks as the lecture topics and language functionality became more complex.
The \textsc{Stella} language evolves thoughout the course from
the core language, which is a simple functional language, to a language with imperative objects~\cite[\S 18]{TaPL} or an ML-style language with universal polymorphism~\cite[\S 23]{TaPL}.
The organization of blocks was as follows:

\subsubsection*{Block 1. Simple Types}

The first block covers the terminology and fundamental concepts necessary to completely understand
simply-typed lambda calculus and implement typechecking for \textsc{Stella} Core, as well as some
simple extensions from the \emph{Syntactic Sugar and Derived Forms}, \emph{Base Types}, and \emph{Simple Types} categories.

The coding assignments in this block require students to implement:
\begin{enumerate}
  \item A typechecker for \textsc{Stella} Core, covering \stella{Nat}, \stella{Bool}, and function types.
  \item A typechecker for simple types: \stella{Unit}, pairs, and sum types.
  \item For extra credit, students may implement a typechecker with support for some of additional extensions:
  variants, tuples, records, \stella{let}-bindings, \stella{letrec}-bindings, nested pattern matching (without exhaustiveness checks), type aliasing, and general recursion.
\end{enumerate}

A theoretical assignment is issued, focusing mainly on sum types and pattern matching,
making sure, students properly understand these concepts.

\subsubsection*{Block 2. Normalization and Recursive Types}

The second block covers theoretical topics,
discussing normalization properties of simply typed lambda calculus~\cite[\S 12]{TaPL}
and talking about recursive types~\cite[\S 20]{TaPL}.
No new assignments are issued in this block, since students have an active assignment from the previous block.

\subsubsection*{Block 3. Imperative Objects}

In the third block, we focus on adding imperative features like references~\cite[\S 13]{TaPL} and exceptions~\cite[\S 14]{TaPL},
and also discuss subtyping, aiming to arrive at a language that supports imperative objects~\cite[\S 18]{TaPL}.
We also discuss Featherweight Java~\cite{IgarashiPierceWadler2001} and Welterweight Java~\cite{OstlundWrigstad2010}.
The latter extends Featherweight Java with imperative, stateful features, as well as thread-based and lock-based concurrency.
While \textsc{Stella} does not support nominal types, students are still asked to complete theoretical excercises using Welterweight Java.

The coding assignment in this block requires students to implement:
\begin{enumerate}
  \item A typechecker supporting sequencing, mutable references, unrecoverable errors, and records with subtyping.
  \item For extra credit, students may implement a typechecker with support for some of the additional extensions:
  subtyping for variants, exceptions with a fixed (super)type, \stella{Top} and \stella{Bot} types, and exceptions with an open variant type.
\end{enumerate}

\subsubsection*{Block 4. Type Reconstruction and Universal Types}

The fourth block focuses on \emph{type reconstruction}~\cite[\S 22]{TaPL} (aka \emph{type inference}) and universal types~\cite[\S 23]{TaPL}.
Additionally, we discuss Hindley-Milder type system~\cite{Hindley1969,Milner1978} and the type inference algorithm for it~\cite{Damas1984}.

The coding assignment in this block requires students to implement type variables,
universal types, and generic function declarations.
The implementation of the type checking here is complicated since students have
to take care with type variables bound by the \stella{forall} quantifier in the polymorphic types.


\subsubsection*{Block 5. Runtime for Lazy Functional Languages}

This last block is dedicated to the exploration of runtime for lazy functional languages.
While this does not directly correspond to type checking, it helps students appreciate
sum types and variants more, as they see how they can be efficiently implemented in a language like Haskell.

\subsection{Assessment}

Once a week at the start of the lecture, students complete a brief quiz consisting of simple theoretical questions (typically, multiple-choice or matching questions).
The quiz is intended to be finished in the classroom within 10 minutes and serves mostly as a tool to facilitate attendance and wake students before the lecture.

To assess the students more comprehensively, three theoretical problem sets are prepared and assigned as homework for one week each.
In the labs, students are working on their coding exercises and receive help from teaching assistants.
Most assignments have an extra credit part.

Students' grades are computed based on completion of assignments and quizzes.
In case of late submissions or to allow for an improvement of their grade,
students are allowed to attend an oral exam to defend their implementation and answer some more theoretical questions
for some additional credit.

\subsubsection{Solutions for Assignment 1}

After completing the first coding assignment, we provide students with a reference implementation in C++ and Java.
In our course, these languages turned out to be most popular among the students, but presented particular implementation challenges.
For instance, in C++, students struggled with the complex handling of raw pointers and segfaults, partially due to raw pointers in the generated code from BNFC.
An additional motivation for sharing our solution with the students was that the first assignment is the basis for all subsequent assignments,
and it is important that it has a good structure and the ability to be easily extended.
This allowed the students to compare and refine their implementations or build entirely on ours.
We find that without a reference implementation for the first assignment,
many students would waste time rewriting the basic logic for each assignment,
which is unacceptable given the course's time frame.



\subsubsection{Collecting Student Submissions}

For quizzes, we relied on the Moodle platform provided by Innopolis University and automatically graded using it.
Solutions to theoretical problem sets have been also submitted through Moodle as PDF files, then graded manually.

For coding exercises, we considered using platforms such as Codeforces\footnote{\url{https://codeforces.com/}}.
Unfortunately, such platforms usually deal with a single-module submissions, which cannot accommodate for the complex compiler project structure.

Consequently, we redirected our focus to GitHub Classroom\footnote{\url{https://classroom.github.com/}} due to its advantages,
including the capability to create repositories in various programming languages as templates for students to clone and commence their projects,
as well as to monitor student submissions by tracking commits.
Although we chose GitHub Classroom, it turned out not ideal for the following reasons:
\begin{enumerate}
    \item It is difficult to update the assignment template (e.g. fix problems, add tests).
    \item It is impossible to offer templates in different languages.
    \item Consecutive assignments have separate repositories,
    it is impossible to submit an individual commit as a solution to a particular assignment.
    To start a new assignment, students had to clone the template each time and manually apply their changes from the previous assignment.
\end{enumerate}

In the end, taking into account the aforementioned issues, GitHub Classroom has allowed us to keep track
of students' work throughout the course: tracking their submission times, work completed, and assessments.

\subsection{Results}

The course was conducted for two groups with a total of 49 (active) students.
A final exam, which was optional and offered to all, was taken by approximately 20\% of the class.
To provide students with the ability to obtain additional guidance, make inquiries, and receive notifications,
we provided a Telegram chat room in addition to the Moodle and GitHub Classroom platforms previously mentioned.
Students completed their projects using various programming languages.


Students selected different languages for their implementation. The distribution
of submissions by language is presented in \cref{table-lang-distribution}.
Students have used C++ and Java extensively in the previous courses, so these were the natural choices for most students.
Kotlin was chosen by many students even when they did not know the language before,
due to simpler code structure. In particular, while the Java template featured the Visitor pattern
which many of the students struggled with (despite being exposed to it in other courses),
in Kotlin students were able to use exhaustive pattern matching through \texttt{when}-expressions,
which they found very convenient when dealing with the nodes of the abstract syntax tree.


Students had some freedom in selecting a subset of extensions to implement in their submissions.
In \cref{table-features-in-number}, we show for each feature,
how many students have implemented it fully (passing all tests) or partially.
\emph{Universal Types} extension was not implemented correctly by many students
due to the problem of name captures, which is non-trivial to implement and
also does not occur in small test programs that students used for debugging.
Many students who attempted to implement \emph{Stuctural Patterns}
struggled with handling contexts properly.
Often the reason was due to incorrect handling of mutable state in an implementation of the Visitor pattern.
When students approached the implementation with a functional style, they were less likely to make mistakes.

At the beginning of the course, we have encountered a major problem
when students were not able to work with the \texttt{Makefile}-based templates
that we have prepared originally.
Setting up the environment and getting the project to a point
where they could begin writing code took one week,
which has affected the schedule of the course negatively,
and forced us to skip some assignments.
Having learned from this experience, we plan to enhance IDE-friendly templates
for the future installation of the course.

\cref{table-language-to-grades} presents the distribution of the final grades for the course according to different programming languages.
With a relatively low number of students, it is hard to draw any definitive conclusions from this data.
It is perhaps not so surprising to see many good grades for popular languages like Java, Kotlin, and C++,
since these languages had most support from the course team, templates, and students were free to communicate and help each other in case of technical difficulties.
On the other hand, it is interesting to note that students who selected languages that initially lacked templates for implementation (Swift, Haskell, TypeScript) also achieved good grades.
The students who used these languages are relatively autonomous and proficient, but we also believe that these languages possess features (such as pattern matching) that noticeably
simplify development of certain parts of a typechecker, which might have helped these students as well.

\begin{table}[]
  \begin{center}
  \begin{tabular}{|l|r|}
    \hline
    Programming Language  & Submissions \\ \hline
    C++                   & 17              \\ \hline
    Java                  & 12              \\ \hline
    Kotlin                & 12              \\ \hline
    Python                & 3               \\ \hline
    JavaScript            & 2               \\ \hline
    Haskell               & 1               \\ \hline
    Swift                 & 1               \\ \hline
    TypeScript            & 1               \\ \hline
  \end{tabular}
  \end{center}
  \caption{Language distribution.}
  \label{table-lang-distribution}
\end{table}

\begin{table}[]
  \begin{center}
  \begin{tabular}{|l|r|r|r|r|}
  \hline
      Feature & Full & Partial & Full, \% & Partial, \% \\ \hline
      Core & 49 & 0 & 100\% & 0\% \\ \hline
      Records & 46 & 0 & 94\% & 0\% \\ \hline
      Pairs & 45 & 0 & 92\% & 0\% \\ \hline
      Unit type & 44 & 0 & 90\% & 0\% \\ \hline
      Sequencing & 43 & 0 & 88\% & 0\% \\ \hline
      References & 42 & 0 & 86\% & 0\% \\ \hline
      Sum types & 39 & 0 & 80\% & 0\% \\ \hline
      Errors & 39 & 0 & 80\% & 0\% \\ \hline
      Subtyping for records & 39 & 0 & 80\% & 0\% \\ \hline
      Tuples & 38 & 3 & 78\% & 6\% \\ \hline
      Universal types & 17 & 21 & 35\% & 43\% \\ \hline
      Top and Bot types & 14 & 8 & 29\% & 16\% \\ \hline
      Records & 12 & 5 & 24\% & 10\% \\ \hline
      Exceptions with a fixed type & 11 & 0 & 22\% & 0\% \\ \hline
      Exceptions with an open variant type & 10 & 0 & 20\% & 0\% \\ \hline
      LetRec-binding & 9 & 0 & 18\% & 0\% \\ \hline
      Subtyping for variants & 9 & 0 & 18\% & 0\% \\ \hline
      Variants & 7 & 0 & 14\% & 0\% \\ \hline
      Let-binding & 7 & 2 & 14\% & 4\% \\ \hline
      Type aliases & 6 & 1 & 12\% & 2\% \\ \hline
      Structural patterns & 3 & 21 & 6\% & 43\% \\ \hline
      General recursion & 3 & 20 & 6\% & 41\% \\ \hline
  \end{tabular}
  \end{center}
  \caption{Implemented features in numbers.}
  \label{table-features-in-number}
\end{table}

\begin{table}[]
  \begin{center}
  \begin{tabular}{|l|r|r|r|r|r|r|r|r|}
  \hline
      Language & A & A, \% & B & B, \% & C & C, \% & D & D, \% \\ \hline
      C++ & 13 & 76\% & 1 & 6\% & 3 & 18\% & 0 & 0\% \\ \hline
      Java & 12 & 75\% & 2 & 17\% & 0 & 0\% & 1 & 8\% \\ \hline
      Kotlin & 8 & 67\% & 4 & 33\% & 0 & 0\% & 0 & 0\% \\ \hline
      Python & 1 & 33\% & 1 & 33\% & 0 & 0\% & 1 & 33\% \\ \hline
      JavaScript & 1 & 50\% & 1 & 50\% & 0 & 0\% & 0 & 0\% \\ \hline
      Haskell & 0 & 0\% & 1 & 100\% & 0 & 0\% & 0 & 0\% \\ \hline
      Swift & 1 & 100\% & 0 & 0\% & 0 & 0\% & 0 & 0\% \\ \hline
      TypeScript & 0 & 0\% & 1 & 100\% & 0 & 0\% & 0 & 0\% \\ \hline
  \end{tabular}
  \end{center}
  \caption{Language to grades.}
  \label{table-language-to-grades}
\end{table}

\section{Conclusion and Future Work}
\label{section:conclusion}

We have presented a half-semester course that focuses on
the study and implementation of type systems, supported by a special programming language \textsc{Stella}
designed to facilitate students' learning process.
The language is intended to accompany a well established textbook by Pierce~\cite{TaPL},
and our preliminary experience shows that students that are used to C-like syntax
absorb the textbook material better when implementing \textsc{Stella}, at least
compared to implementing raw typed lambda calculi.

There are still many ways to improve \textsc{Stella} to better accommodate
the needs of the educational process.
Nominal type systems prevail in modern programming languages, so \textsc{Stella}
should support them to enable a more fruitful discussion and comparison with structural types.
Bounded universal quantification is currently not supported in \textsc{Stella},
but we plan to add it, since it allows for a more direct experience with type systems
that incorporate both parametric polymorphism and subtyping, and is used a lot
in languages like Java, Scala, Kotlin, C\#, OCaml.
Annotating functions with the types of possible exceptions is also currently not supported
and provides for a relatively straightforward but useful language extension.
Row polymorphism~\cite{Wand1989} is a relatively rare but important concept, that is currently
not explored by \textsc{Stella}.

Although we focus mostly on type systems, we are also interested in compilation
techniques for functional programming languages. \textsc{Stella}'s extension
system may serve as a basis for studying different compiler backends.
In particular, while in our course we touch on STG~\cite{PeytonJones1992},
other intermediate representations such as GRIN~\cite{BoquistJohnsson1997,Boquist1999} and HVM~\cite{Mazza2007,HVM} deserve attention.
However, more extensions related to operational semantics should be considered
for such an exploration. We feel that this belongs to a separate course.

Another improvement to the course could be to improve the quality and automation of tests.
Instead of relying merely on the exit code of the type checker,
we should also take into account the type error message,
which should follow a standard specified in \textsc{Stella} documentation and available in the playground.
A partial implementation of that idea is given by error tags, which are provided in Stella output. Additionally,
in some situations, the playground is able to suggest some alternative errors, allowing for some variation in student's implementations.
An approach to automation of testing similar to L{\"u}bke, Fuger, Bahnsen, Billerbeck, and Schupp~\cite{LubkeFugerBahnsenBillerbeckSchupp2023}
could also be employed in the future.







\textbf{Acknowledgements.}
We thank Artem Murashko, Timur Iakshigulov, Alexandr Kudasov, Iskander Nafikov, and Danila Korneenko
for their help with Swift, Go, Rust, and Kotlin templates.
We thank Asem Abdelhady for his contribution to the test suite.

\nocite{*}
\bibliographystyle{eptcs}
\bibliography{ms}

\begin{thebibliography}{10}
\providecommand{\bibitemdeclare}[2]{}
\providecommand{\surnamestart}{}
\providecommand{\surnameend}{}
\providecommand{\urlprefix}{Available at }
\providecommand{\url}[1]{\texttt{#1}}
\providecommand{\href}[2]{\texttt{#2}}
\providecommand{\urlalt}[2]{\href{#1}{#2}}
\providecommand{\doi}[1]{doi:\urlalt{https://doi.org/#1}{#1}}
\providecommand{\eprint}[1]{arXiv:\urlalt{https://arxiv.org/abs/#1}{#1}}
\providecommand{\bibinfo}[2]{#2}

\bibitemdeclare{article}{Aiken1996}
\bibitem{Aiken1996}
\bibinfo{author}{Alexander \surnamestart Aiken\surnameend}
  (\bibinfo{year}{1996}): \emph{\bibinfo{title}{Cool: A Portable Project for
  Teaching Compiler Construction}}.
\newblock {\slshape \bibinfo{journal}{SIGPLAN Not.}}
  \bibinfo{volume}{31}(\bibinfo{number}{7}), p. \bibinfo{pages}{19–24},
  \doi{10.1145/381841.381847}.

\bibitemdeclare{book}{Appel2004ML}
\bibitem{Appel2004ML}
\bibinfo{author}{Andrew~W. \surnamestart Appel\surnameend}
  (\bibinfo{year}{2004}): \emph{\bibinfo{title}{Modern Compiler Implementation
  in ML}}.
\newblock \bibinfo{publisher}{Cambridge University Press},
  \bibinfo{address}{USA}.

\bibitemdeclare{book}{AppelGinsburg2004C}
\bibitem{AppelGinsburg2004C}
\bibinfo{author}{Andrew~W. \surnamestart Appel\surnameend} \&
  \bibinfo{author}{Maia \surnamestart Ginsburg\surnameend}
  (\bibinfo{year}{2004}): \emph{\bibinfo{title}{Modern Compiler Implementation
  in C}}.
\newblock \bibinfo{publisher}{Cambridge University Press},
  \bibinfo{address}{USA}.

\bibitemdeclare{book}{AppelPalsberg2003Java}
\bibitem{AppelPalsberg2003Java}
\bibinfo{author}{Andrew~W. \surnamestart Appel\surnameend} \&
  \bibinfo{author}{Jens \surnamestart Palsberg\surnameend}
  (\bibinfo{year}{2003}): \emph{\bibinfo{title}{Modern Compiler Implementation
  in Java}}, \bibinfo{edition}{2nd} edition.
\newblock \bibinfo{publisher}{Cambridge University Press},
  \bibinfo{address}{USA}.

\bibitemdeclare{inproceedings}{Appelbe1979}
\bibitem{Appelbe1979}
\bibinfo{author}{Bill \surnamestart Appelbe\surnameend} (\bibinfo{year}{1979}):
  \emph{\bibinfo{title}{Teaching Compiler Development}}.
\newblock In: {\slshape \bibinfo{booktitle}{Proceedings of the Tenth SIGCSE
  Technical Symposium on Computer Science Education}}, \bibinfo{series}{SIGCSE
  '79}, \bibinfo{publisher}{Association for Computing Machinery},
  \bibinfo{address}{New York, NY, USA}, p. \bibinfo{pages}{23–27},
  \doi{10.1145/800126.809546}.

\bibitemdeclare{inproceedings}{BerezunBoulytchev2022}
\bibitem{BerezunBoulytchev2022}
\bibinfo{author}{Daniil \surnamestart Berezun\surnameend} \&
  \bibinfo{author}{Dmitry \surnamestart Boulytchev\surnameend}
  (\bibinfo{year}{2022}): \emph{\bibinfo{title}{Reimplementing the Wheel:
  Teaching Compilers with a Small Self-Contained One}}.
\newblock In \bibinfo{editor}{Peter \surnamestart Achten\surnameend} \&
  \bibinfo{editor}{Elena \surnamestart Machkasova\surnameend}, editors:
  {\slshape \bibinfo{booktitle}{Proceedings Tenth and Eleventh International
  Workshop on Trends in Functional Programming In Education, {TFPIE} 2021 /
  2022, Krak{\'{o}}w, Poland (online), 16th February 2021 / 16th March 2022}},
  {\slshape \bibinfo{series}{{EPTCS}}} \bibinfo{volume}{363}, pp.
  \bibinfo{pages}{22--43}, \doi{10.4204/EPTCS.363.2}.

\bibitemdeclare{phdthesis}{Boquist1999}
\bibitem{Boquist1999}
\bibinfo{author}{Urban \surnamestart Boquist\surnameend}
  (\bibinfo{year}{1999}): \emph{\bibinfo{title}{Code optimization techniques
  for lazy functional languages}}.
\newblock Ph.D. thesis, \bibinfo{school}{Chalmers Tekniska H{\"o}gskola}.

\bibitemdeclare{inproceedings}{BoquistJohnsson1997}
\bibitem{BoquistJohnsson1997}
\bibinfo{author}{Urban \surnamestart Boquist\surnameend} \&
  \bibinfo{author}{Thomas \surnamestart Johnsson\surnameend}
  (\bibinfo{year}{1997}): \emph{\bibinfo{title}{The GRIN project: A highly
  optimising back end for lazy functional languages}}.
\newblock In: {\slshape \bibinfo{booktitle}{Implementation of Functional
  Languages: 8th International Workshop, IFL'96 Bad Godesberg, Germany,
  September 16--18, 1996 Selected Papers 8}}, \bibinfo{organization}{Springer},
  pp. \bibinfo{pages}{58--84}, \doi{10.1007/3-540-63237-9_19}.

\bibitemdeclare{book}{Brady2017}
\bibitem{Brady2017}
\bibinfo{author}{Edwin \surnamestart Brady\surnameend} (\bibinfo{year}{2017}):
  \emph{\bibinfo{title}{Type-driven development with Idris}}.
\newblock \bibinfo{publisher}{Simon and Schuster}.

\bibitemdeclare{article}{CoquandHuet1988}
\bibitem{CoquandHuet1988}
\bibinfo{author}{Thierry \surnamestart Coquand\surnameend} \&
  \bibinfo{author}{Gérard \surnamestart Huet\surnameend}
  (\bibinfo{year}{1988}): \emph{\bibinfo{title}{The calculus of
  constructions}}.
\newblock {\slshape \bibinfo{journal}{Information and Computation}}
  \bibinfo{volume}{76}(\bibinfo{number}{2}), pp. \bibinfo{pages}{95--120},
  \doi{10.1016/0890-5401(88)90005-3}.

\bibitemdeclare{phdthesis}{Damas1984}
\bibitem{Damas1984}
\bibinfo{author}{Lu{\'{\i}}s \surnamestart Damas\surnameend}
  (\bibinfo{year}{1984}): \emph{\bibinfo{title}{Type assignment in programming
  languages}}.
\newblock Ph.D. thesis, \bibinfo{school}{University of Edinburgh, {UK}}.
\newblock \urlprefix\url{https://hdl.handle.net/1842/13555}.

\bibitemdeclare{inproceedings}{EysholdtBehrens2010}
\bibitem{EysholdtBehrens2010}
\bibinfo{author}{Moritz \surnamestart Eysholdt\surnameend} \&
  \bibinfo{author}{Heiko \surnamestart Behrens\surnameend}
  (\bibinfo{year}{2010}): \emph{\bibinfo{title}{Xtext: implement your language
  faster than the quick and dirty way}}.
\newblock In: {\slshape \bibinfo{booktitle}{Proceedings of the ACM
  international conference companion on Object oriented programming systems
  languages and applications companion}}, pp. \bibinfo{pages}{307--309},
  \doi{10.1145/1869542.1869625}.

\bibitemdeclare{inproceedings}{BNFC}
\bibitem{BNFC}
\bibinfo{author}{Markus \surnamestart Forsberg\surnameend} \&
  \bibinfo{author}{Aarne \surnamestart Ranta\surnameend}
  (\bibinfo{year}{2004}): \emph{\bibinfo{title}{BNF Converter}}.
\newblock In: {\slshape \bibinfo{booktitle}{Proceedings of the 2004 ACM SIGPLAN
  Workshop on Haskell}}, \bibinfo{series}{Haskell '04},
  \bibinfo{publisher}{Association for Computing Machinery},
  \bibinfo{address}{New York, NY, USA}, p. \bibinfo{pages}{94–95},
  \doi{10.1145/1017472.1017475}.

\bibitemdeclare{article}{Girard1986}
\bibitem{Girard1986}
\bibinfo{author}{Jean-Yves \surnamestart Girard\surnameend}
  (\bibinfo{year}{1986}): \emph{\bibinfo{title}{The system F of variable types,
  fifteen years later}}.
\newblock {\slshape \bibinfo{journal}{Theoretical Computer Science}}
  \bibinfo{volume}{45}, pp. \bibinfo{pages}{159--192},
  \doi{10.1016/0304-3975(86)90044-7}.

\bibitemdeclare{article}{Hindley1969}
\bibitem{Hindley1969}
\bibinfo{author}{R.~\surnamestart Hindley\surnameend} (\bibinfo{year}{1969}):
  \emph{\bibinfo{title}{The Principal Type-Scheme of an Object in Combinatory
  Logic}}.
\newblock {\slshape \bibinfo{journal}{Transactions of the American Mathematical
  Society}} \bibinfo{volume}{146}, pp. \bibinfo{pages}{29--60},
  \doi{10.1090/S0002-9947-1969-0253905-6}.
\newblock \urlprefix\url{http://www.jstor.org/stable/1995158}.

\bibitemdeclare{article}{IgarashiPierceWadler2001}
\bibitem{IgarashiPierceWadler2001}
\bibinfo{author}{Atsushi \surnamestart Igarashi\surnameend},
  \bibinfo{author}{Benjamin~C. \surnamestart Pierce\surnameend} \&
  \bibinfo{author}{Philip \surnamestart Wadler\surnameend}
  (\bibinfo{year}{2001}): \emph{\bibinfo{title}{Featherweight Java: A Minimal
  Core Calculus for Java and GJ}}.
\newblock {\slshape \bibinfo{journal}{ACM Trans. Program. Lang. Syst.}}
  \bibinfo{volume}{23}(\bibinfo{number}{3}), p. \bibinfo{pages}{396–450},
  \doi{10.1145/503502.503505}.

\bibitemdeclare{article}{PeytonJones1992}
\bibitem{PeytonJones1992}
\bibinfo{author}{Simon L.~Peyton \surnamestart Jones\surnameend}
  (\bibinfo{year}{1992}): \emph{\bibinfo{title}{Implementing Lazy Functional
  Languages on Stock Hardware: The Spineless Tagless G-Machine}}.
\newblock {\slshape \bibinfo{journal}{J. Funct. Program.}}
  \bibinfo{volume}{2}(\bibinfo{number}{2}), pp. \bibinfo{pages}{127--202},
  \doi{10.1017/S0956796800000319}.

\bibitemdeclare{inproceedings}{Krishnaswami2009}
\bibitem{Krishnaswami2009}
\bibinfo{author}{Neelakantan~R. \surnamestart Krishnaswami\surnameend}
  (\bibinfo{year}{2009}): \emph{\bibinfo{title}{Focusing on pattern matching}}.
\newblock In \bibinfo{editor}{Zhong \surnamestart Shao\surnameend} \&
  \bibinfo{editor}{Benjamin~C. \surnamestart Pierce\surnameend}, editors:
  {\slshape \bibinfo{booktitle}{Proceedings of the 36th {ACM} {SIGPLAN-SIGACT}
  Symposium on Principles of Programming Languages, {POPL} 2009, Savannah, GA,
  USA, January 21-23, 2009}}, \bibinfo{publisher}{{ACM}}, pp.
  \bibinfo{pages}{366--378}, \doi{10.1145/1480881.1480927}.

\bibitemdeclare{inproceedings}{LubkeFugerBahnsenBillerbeckSchupp2023}
\bibitem{LubkeFugerBahnsenBillerbeckSchupp2023}
\bibinfo{author}{Ole \surnamestart L{\"u}bke\surnameend},
  \bibinfo{author}{Konrad \surnamestart Fuger\surnameend},
  \bibinfo{author}{Fin~Hendrik \surnamestart Bahnsen\surnameend},
  \bibinfo{author}{Katrin \surnamestart Billerbeck\surnameend} \&
  \bibinfo{author}{Sibylle \surnamestart Schupp\surnameend}
  (\bibinfo{year}{2023}): \emph{\bibinfo{title}{How to Derive an Electronic
  Functional Programming Exam from a Paper Exam with Proofs and Programming
  Tasks}}.
\newblock In: {\slshape \bibinfo{booktitle}{Trends in Functional Programming in
  Education (TFPIE)}}.

\bibitemdeclare{misc}{HVM}
\bibitem{HVM}
\bibinfo{author}{Victor \surnamestart Maia\surnameend} (\bibinfo{year}{2023}):
  \emph{\bibinfo{title}{Higher-order Virtual Machine (HVM)}}.
\newblock \urlprefix\url{https://github.com/HigherOrderCO/hvm}.

\bibitemdeclare{article}{Maranget2007}
\bibitem{Maranget2007}
\bibinfo{author}{Luc \surnamestart Maranget\surnameend} (\bibinfo{year}{2007}):
  \emph{\bibinfo{title}{Warnings for pattern matching}}.
\newblock {\slshape \bibinfo{journal}{Journal of Functional Programming}}
  \bibinfo{volume}{17}(\bibinfo{number}{3}), p. \bibinfo{pages}{387–421},
  \doi{10.1017/S0956796807006223}.

\bibitemdeclare{book}{MartinLof1984}
\bibitem{MartinLof1984}
\bibinfo{author}{Per \surnamestart Martin-L{\"o}f\surnameend} \&
  \bibinfo{author}{Giovanni \surnamestart Sambin\surnameend}
  (\bibinfo{year}{1984}): \emph{\bibinfo{title}{Intuitionistic type theory}}.
\newblock \bibinfo{volume}{9}, \bibinfo{publisher}{Bibliopolis Naples}.

\bibitemdeclare{inproceedings}{MatsakisKlock2014}
\bibitem{MatsakisKlock2014}
\bibinfo{author}{Nicholas~D. \surnamestart Matsakis\surnameend} \&
  \bibinfo{author}{Felix~S. \surnamestart Klock\surnameend}
  (\bibinfo{year}{2014}): \emph{\bibinfo{title}{The Rust Language}}.
\newblock In: {\slshape \bibinfo{booktitle}{Proceedings of the 2014 ACM SIGAda
  Annual Conference on High Integrity Language Technology}},
  \bibinfo{series}{HILT '14}, \bibinfo{publisher}{Association for Computing
  Machinery}, \bibinfo{address}{New York, NY, USA}, p.
  \bibinfo{pages}{103–104}, \doi{10.1145/2663171.2663188}.

\bibitemdeclare{article}{Mazza2007}
\bibitem{Mazza2007}
\bibinfo{author}{Damiano \surnamestart Mazza\surnameend}
  (\bibinfo{year}{2007}): \emph{\bibinfo{title}{A denotational semantics for
  the symmetric interaction combinators}}.
\newblock {\slshape \bibinfo{journal}{Mathematical Structures in Computer
  Science}} \bibinfo{volume}{17}(\bibinfo{number}{3}), pp.
  \bibinfo{pages}{527--562}, \doi{10.1017/S0960129507006135}.

\bibitemdeclare{article}{Mernik2003}
\bibitem{Mernik2003}
\bibinfo{author}{M.~\surnamestart Mernik\surnameend} \&
  \bibinfo{author}{V.~\surnamestart Zumer\surnameend} (\bibinfo{year}{2003}):
  \emph{\bibinfo{title}{An educational tool for teaching compiler
  construction}}.
\newblock {\slshape \bibinfo{journal}{IEEE Transactions on Education}}
  \bibinfo{volume}{46}(\bibinfo{number}{1}), pp. \bibinfo{pages}{61--68},
  \doi{10.1109/TE.2002.808277}.

\bibitemdeclare{article}{Milner1978}
\bibitem{Milner1978}
\bibinfo{author}{Robin \surnamestart Milner\surnameend} (\bibinfo{year}{1978}):
  \emph{\bibinfo{title}{A theory of type polymorphism in programming}}.
\newblock {\slshape \bibinfo{journal}{Journal of Computer and System Sciences}}
  \bibinfo{volume}{17}(\bibinfo{number}{3}), pp. \bibinfo{pages}{348--375},
  \doi{10.1016/0022-0000(78)90014-4}.

\bibitemdeclare{book}{Muchnick1998}
\bibitem{Muchnick1998}
\bibinfo{author}{Steven~S. \surnamestart Muchnick\surnameend}
  (\bibinfo{year}{1998}): \emph{\bibinfo{title}{Advanced Compiler Design and
  Implementation}}.
\newblock \bibinfo{publisher}{Morgan Kaufmann Publishers Inc.},
  \bibinfo{address}{San Francisco, CA, USA}.

\bibitemdeclare{misc}{GHCJS}
\bibitem{GHCJS}
\bibinfo{author}{Victor \surnamestart Nazarov\surnameend},
  \bibinfo{author}{Hamish \surnamestart Mackenzie\surnameend} \&
  \bibinfo{author}{Luite \surnamestart Stegeman\surnameend}
  (\bibinfo{year}{2015}): \emph{\bibinfo{title}{GHCJS Haskell to JavaScript
  compiler}}.
\newblock \urlprefix\url{https://github.com/ghcjs/ghcjs}.

\bibitemdeclare{article}{OrtinZapicoCueva2007}
\bibitem{OrtinZapicoCueva2007}
\bibinfo{author}{Francisco \surnamestart Ortin\surnameend},
  \bibinfo{author}{Daniel \surnamestart Zapico\surnameend} \&
  \bibinfo{author}{Juan~Manuel \surnamestart Cueva\surnameend}
  (\bibinfo{year}{2007}): \emph{\bibinfo{title}{Design Patterns for Teaching
  Type Checking in a Compiler Construction Course}}.
\newblock {\slshape \bibinfo{journal}{IEEE Transactions on Education}}
  \bibinfo{volume}{50}(\bibinfo{number}{3}), pp. \bibinfo{pages}{273--283},
  \doi{10.1109/TE.2007.901983}.

\bibitemdeclare{inproceedings}{OstlundWrigstad2010}
\bibitem{OstlundWrigstad2010}
\bibinfo{author}{Johan \surnamestart \"{O}stlund\surnameend} \&
  \bibinfo{author}{Tobias \surnamestart Wrigstad\surnameend}
  (\bibinfo{year}{2010}): \emph{\bibinfo{title}{Welterweight Java}}.
\newblock In: {\slshape \bibinfo{booktitle}{Proceedings of the 48th
  International Conference on Objects, Models, Components, Patterns}},
  \bibinfo{series}{TOOLS'10}, \bibinfo{publisher}{Springer-Verlag},
  \bibinfo{address}{Berlin, Heidelberg}, p. \bibinfo{pages}{97–116},
  \doi{10.1007/978-3-642-13953-6_6}.

\bibitemdeclare{book}{TaPL}
\bibitem{TaPL}
\bibinfo{author}{Benjamin~C. \surnamestart Pierce\surnameend}
  (\bibinfo{year}{2002}): \emph{\bibinfo{title}{Types and programming
  languages}}.
\newblock \bibinfo{publisher}{{MIT} Press}.

\bibitemdeclare{article}{Plotkin1977LCFCA}
\bibitem{Plotkin1977LCFCA}
\bibinfo{author}{Gordon~D. \surnamestart Plotkin\surnameend}
  (\bibinfo{year}{1977}): \emph{\bibinfo{title}{LCF Considered as a Programming
  Language}}.
\newblock {\slshape \bibinfo{journal}{Theor. Comput. Sci.}}
  \bibinfo{volume}{5}, pp. \bibinfo{pages}{223--255},
  \doi{10.1016/0304-3975(77)90044-5}.
\newblock \urlprefix\url{https://api.semanticscholar.org/CorpusID:53785015}.

\bibitemdeclare{inproceedings}{Reynolds1974}
\bibitem{Reynolds1974}
\bibinfo{author}{John~C \surnamestart Reynolds\surnameend}
  (\bibinfo{year}{1974}): \emph{\bibinfo{title}{Towards a theory of type
  structure}}.
\newblock In: {\slshape \bibinfo{booktitle}{Programming Symposium: Proceedings,
  Colloque sur la Programmation Paris, April 9--11, 1974}},
  \bibinfo{organization}{Springer}, pp. \bibinfo{pages}{408--425},
  \doi{10.1007/3-540-06859-7_148}.

\bibitemdeclare{inproceedings}{Sestoft1996}
\bibitem{Sestoft1996}
\bibinfo{author}{Peter \surnamestart Sestoft\surnameend}
  (\bibinfo{year}{1996}): \emph{\bibinfo{title}{{ML} Pattern Match Compilation
  and Partial Evaluation}}.
\newblock In \bibinfo{editor}{Olivier \surnamestart Danvy\surnameend},
  \bibinfo{editor}{Robert \surnamestart Gl{\"{u}}ck\surnameend} \&
  \bibinfo{editor}{Peter \surnamestart Thiemann\surnameend}, editors: {\slshape
  \bibinfo{booktitle}{Partial Evaluation, International Seminar, Dagstuhl
  Castle, Germany, February 12-16, 1996, Selected Papers}}, {\slshape
  \bibinfo{series}{Lecture Notes in Computer Science}} \bibinfo{volume}{1110},
  \bibinfo{publisher}{Springer}, pp. \bibinfo{pages}{446--464},
  \doi{10.1007/3-540-61580-6\_22}.

\bibitemdeclare{inproceedings}{Wand1989}
\bibitem{Wand1989}
\bibinfo{author}{M.~\surnamestart Wand\surnameend} (\bibinfo{year}{1989}):
  \emph{\bibinfo{title}{Type inference for record concatenation and multiple
  inheritance}}.
\newblock In: {\slshape \bibinfo{booktitle}{[1989] Proceedings. Fourth Annual
  Symposium on Logic in Computer Science}}, pp. \bibinfo{pages}{92--97},
  \doi{10.1109/LICS.1989.39162}.

\bibitemdeclare{misc}{DifferentiableSwift}
\bibitem{DifferentiableSwift}
\bibinfo{author}{Richard \surnamestart Wei\surnameend}, \bibinfo{author}{Dan
  \surnamestart Zheng\surnameend}, \bibinfo{author}{Marc \surnamestart
  Rasi\surnameend} \& \bibinfo{author}{Bart \surnamestart Chrzaszcz\surnameend}
  (\bibinfo{year}{2023}): \emph{\bibinfo{title}{Differentiable Programming
  Manifesto}}.
\newblock
  \urlprefix\url{https://github.com/apple/swift/blob/main/docs/DifferentiableProgramming.md}.

\bibitemdeclare{book}{Wirth1996}
\bibitem{Wirth1996}
\bibinfo{author}{Niklaus \surnamestart Wirth\surnameend}
  (\bibinfo{year}{1996}): \emph{\bibinfo{title}{Compiler Construction}}.
\newblock \bibinfo{publisher}{Addison Wesley Longman Publishing Co., Inc.},
  \bibinfo{address}{USA}.

\bibitemdeclare{article}{Yorgey2023}
\bibitem{Yorgey2023}
\bibinfo{author}{Brent~A. \surnamestart Yorgey\surnameend}
  (\bibinfo{year}{2023}): \emph{\bibinfo{title}{Disco: A Functional Programming
  Language for Discrete Mathematics}}.
\newblock {\slshape \bibinfo{journal}{Electronic Proceedings in Theoretical
  Computer Science}} \bibinfo{volume}{382}, p. \bibinfo{pages}{64–81},
  \doi{10.4204/eptcs.382.4}.

\end{thebibliography}
\end{document}